\documentclass[%
 reprint,
superscriptaddress,
 amsmath,amssymb,
 aps,prx
]{revtex4-2}

\usepackage{graphicx}
\usepackage{dcolumn}
\usepackage{bm}
\usepackage[colorlinks=true,linkcolor=blue,urlcolor=blue,citecolor=blue]{hyperref}

\usepackage{subfigure}
\usepackage{xcolor}
\usepackage{amsmath}
 \usepackage{physics}
\usepackage{float}
\begin{document}

\preprint{APS/123-QED}
\title{Sensing Single Photon in a Cat State}

\author{Arman}
\email{a19rs004@iiserkol.ac.in}
\affiliation{Department of Physical Sciences, Indian Institute of Science Education and Research Kolkata, Mohanpur-741246, West Bengal, India}

 \author{Gargi Tyagi}
 \email{gargityagi150@gmail.com}
\affiliation{Institute of Applied Sciences, Amity University, Sector-125, Noida-201303, Uttar Pradesh, India} 

\author{Prasanta K. Panigrahi}%
\email{pprasanta@iiserkol.ac.in}
\affiliation{Department of Physical Sciences, Indian Institute of Science Education and Research Kolkata, Mohanpur-741246, West Bengal, India}

\date{\today}

\date{\today}

\begin{abstract}

The cat state is  shown to `store' a single photon through the superposition of its orthogonal counterpart with itself, and an excited oscillator state. Photon addition leads to a $\pi $ phase shift at origin in the observed phase space interference of the Wigner function, which also displays negativity, controlled by the average photon number ($|\alpha|^2$) of coherent states comprising the cat state. The maxima and minima of the sub-Planck tiles in the phase space of the kitten state are interchanged after photon addition, leading to their orthogonality. Interestingly, photon addition to Yurke-Stoler state characterized by Poissonian statistics leads to a sub-Poissonian distribution. 

\end{abstract}

\maketitle


\section{Introduction}

Coherent states of light are classical in nature \cite{sudercoher,glubercoher}, represented by localized Gaussian wave packets with a photon distribution having Poissonian character \cite{ag,fox,gerry}. The center of the wave-packet performs harmonic motion under the temporal evolution of the underlying number states. Remarkably, addition of a single photon has been shown to alter its properties in a measurable way \cite{agarwal,photopluscho2020}, which has found experimental verification \cite{zavatta}. Addition of more photons naturally takes the state closer to number states. The cat state, a superposition of two coherent states $\,(\ket{\alpha} \pm \ket{-\alpha}\,$ with  $\hat{a} \ket{\alpha}=\alpha \ket{\alpha}$,  $\hat{a}$ being the annihilation operator), is well-known for its efficacy in capturing the foundational aspects of quantum mechanics, particularly in the context of measurement induced collapse of the wave function \cite{schrodinger}.

The manifestation of quantumness in cat-state is `minimal' in the sense that even and odd cat-states  are respectively composed of the even and odd number states in the oscillator basis \cite{ag}. Remarkably, the parity difference of even and odd cat states have found profound use in checking the deviation of the original state due to loss of a single photon, enabling error correction through an ancila. The cat state, being superposition of coherent states, is an eigen state of the operator $\hat{a}^2$, and hence reverts back to its original form after loss of two photons \cite{mic,ancilla-steven}. This has led to the so-called cat-code, a tool for continuous variable quantum information processing,  which is robust against photon loss \cite{jiang}. Continuous variable \cite{Continu-vari} macroscopic quantum states are now being experimentally realized with high fidelity in optical platforms \cite{bell, and}. Tuning between discrete and continuous variable measurements have also been made experimentally possible by weak-field homodyne detection \cite{bell2}. Very recently, four component cat state has been  realized in optical  \cite{and2} and microwave regimes \cite{mic}. Optical systems have also been used to produce cluster state, useful for measurement based quantum computing  \cite{anderson, asav}.

The cat state is inherently non-local, being superposition of two displaced Gaussian functions, in  a coordinate space representation. It therefore, is of deep interest to investigate how a single photon is \textit{stored} in this non-local configuration. The imprints of the number and phase information on this non-local state are of  physical interest, as the corresponding  operators are canonical conjugates of each other \cite{pegg}. Here, we study the single photon-added cat state (SPACS) in the phase space to investigate the precise variations in the Wigner distribution. In particular, the signature of the number and phase information on  the cat state is probed through changes in the phase-space interference, made manifest through the Wigner function \cite{wheeler}. Zurek \cite{zurek} and Vogel \cite{vogel} have earlier pointed out the effectiveness of cat and the more general kitten states in precision metrology, making use of the phase-space interference. A small change in any of the two quadratures, conveniently represented as co-ordinate and momentum, makes these states orthogonal and hence distinguishable. This arises from the sub-Fourier sensitivity of the cat state and sub-Planck structures in the Wigner function $W(x,p)$ of the kitten state \cite{praxemayer}. The kitten state, localized in both co-ordinate and momentum space, shows much richer phase-space interference structure, leading to simultaneous sensitivity to co-ordinate and momentum displacements \cite{utpalgosh,zurek2}.

The paper is organised as follows: we start with the photon added coherent state (CS), highlighting the phase space structure resulting from photon addition. We then investigate the measurable effects of photon addition to the even cat state. Our study manifestly reveals duality of the number and phase information in the phase space of the cat state. The addition of a single photon is found to result in the fringe shift by one unit in the phase-space and making the Wigner function negative at the origin. This gets further magnified as a function of parameter ($|\alpha|^2$).

We then study the effect of photon addition on the kitten state. The sub-Planck structure in phase space \cite{olive} is shown to be measurably shifted diagonally due to one photon addition. This leads to interchange of maxima and minima positions, making the states orthogonal and hence distinguishable. The effect of photon addition also affects the statistics. The statistics of the Yurke-Stoler (YS) state is shown to change from Poissonian to sub-Poissonian with addition of one photon. 
As mentioned earlier, the addition of a single photon to the CS leads to measurable changes in its wave function: 

$$\psi_{\alpha}(x) =\psi_0(x-\alpha_1)e^{i\alpha_{2}(x-\alpha_{1}/2)}$$
\begin{equation}
  =\exp{-\frac{(x-\alpha_{1})^2}{2}}e^{i\alpha_{2}(x-\alpha_{1}/2)}.   
\end{equation}

The photon added coherent state is a superposition of the mutually orthogonal displaced first excited and ground state \cite{agarwal}: 
$$ \psi_{(\alpha,1)}(x)  =\left[\psi_{1}(x-\alpha_{1})+\frac{\alpha^{*}}{\sqrt{2}}\psi_{0}(x-\alpha_{1})\right]e^{i\alpha_{2}(x-\alpha_{1}/2)}$$

with $$\psi_{1}(x)=\sqrt{2}x \exp{-\frac{x^2}{2}}.\,\,$$

The resulting negativity of the Wigner function makes it clearly discernible from the coherent state with a positive  Wigner function \cite{parigi, lee}.

\section{Photon added cat state}

The cat state, being a superposition of coherent states, offers the possibility of both number and phase estimation. As is well-known, the number and phase degrees of freedom, being canonically conjugate, obey the quadrature uncertainty and hence will have distinct signature on cat state.

The general cat state is of the form:

\begin{multline} 
   \label{cat0}
\psi_{cat}(x)  =N\left[e^{-(x-\alpha_{1})^{2}/2}e^{i\alpha_{2}(x-\alpha_{1}/2)}\right.\\
+ \left.e^{i\theta}e^{-(x+\alpha_{1})^{2}/2}e^{-i\alpha_{2}(x+\alpha_{1}/2)}\right],
\end{multline}
where 
$$N= \frac{1}{\sqrt{2\sqrt{\pi}(1+ e^{-|\alpha|^{2}}\cos\theta)}}.$$

Evidently, the even and odd cat states \cite{dodonov}, and Yurke-Stoler state \cite{yurke} arise for $\theta=0, \pi \,\textnormal{and}\, \pi/2\,$ respectively.

Loss of one photon changes the even (odd) cat state to the orthogonal odd (even) cat state, leading to a phase flip \cite{mic}:  

\begin{equation}
  a(\ket{\alpha}\pm\ket{-\alpha})=\alpha(\ket{\alpha}\mp\ket{-\alpha}).
\end{equation}

Th cat code uses these states as qubits and is robust against photon loss, as  
two photon loss brings the state  back to its original form. 

The single photon added even cat state carries the number and phase information:
\begin{multline}\label{cat}
\psi_{(cat_{e},1)}(x)  =  \left[\sqrt{2}(x-\alpha_{1})e^{-(x-\alpha_{1})^{2}/2}e^{i\alpha_{2}(x-\alpha_{1}/2)}\right.\\
 +\frac{\alpha*}{\sqrt{2}}e^{-(x-\alpha_{1})^{2}/2}e^{i\alpha_{2}(x-\alpha_{1}/2)}\\
 +\left.\sqrt{2}(x+\alpha_{1})e^{-(x+\alpha_{1})^{2}/2}e^{-i\alpha_{2}(x+\alpha_{1}/2)}\right.\\
 -\left.\frac{\alpha*}{\sqrt{2}}e^{-(x+\alpha_{1})^{2}/2}e^{-i\alpha_{2}(x+\alpha_{1}/2)}\right].
  \end{multline}
Akin  to  the  coherent  state,  it  also  has  the  first  excited state,  leading  to  vanishing of the  wave  function  at  origin and negativity in the Wigner function.  
As will be explicitly shown, the inherent quantumness of the cat state manifests through the phase space interference, which incorporates the phase information due to photon addition.

The Wigner function can be written as sum of four terms, representing the terms responsible for interference effect in phase space and the ones showing particle localization.

Explicitly, 
 $$W_{(cat,1)}(x,p) = \frac{W_{++}+ W_{--} + (W_{+-}+W_{-+})}{\sqrt{\pi}\left((2+|\alpha|^{2})+(2-|\alpha|^{2})e^{-|\alpha|^{2}}\right)}, $$
where 

\begin{multline}
W_{\pm\pm}= \sqrt{\pi}\left(4\left[x(x\mp\alpha_{1})+p(p\mp\alpha_{2})\right]-2+|\alpha|^{2})\right.\\
\times e^{-((x\mp\alpha_{1})^{2}+(p\mp\alpha_{2})^{2})} \end{multline}

represent localization and squeezing of coherent states with the photon addition. 

The phase space interference  of one photon added cat state originates from,
$$ W_{+-}+W_{-+} = 4\sqrt{\pi}\left[2\left(\alpha_{2}x-p\alpha_{1}\right)\sin{2(\alpha_{2}x-\alpha_{1}p)}\right.+$$
\begin{equation}
\left.\left\{2\left(x^{2}+p^{2}\right)-1-|\alpha|^{2}/2\right\}\cos{2(\alpha_{2}x-\alpha_{1}p)}\right]e^{-(x^{2}+p^{2})}.    
\end{equation}

The coefficient of cosine term quantifies the depth of negativity of the Wigner function given at the origin by $- \left(\frac{4+2|\alpha|^2}{\left((2+|\alpha|^{2})+(2-|\alpha|^{2})e^{-|\alpha|^{2}}\right)}\right) $. Evidently, it is controlled by the  parameter $|\alpha|^2$ of the cat state itself.

\onecolumngrid 
\begin{center}
\begin{figure}[H]
\includegraphics[width=19.0 cm,height=10.0 cm]{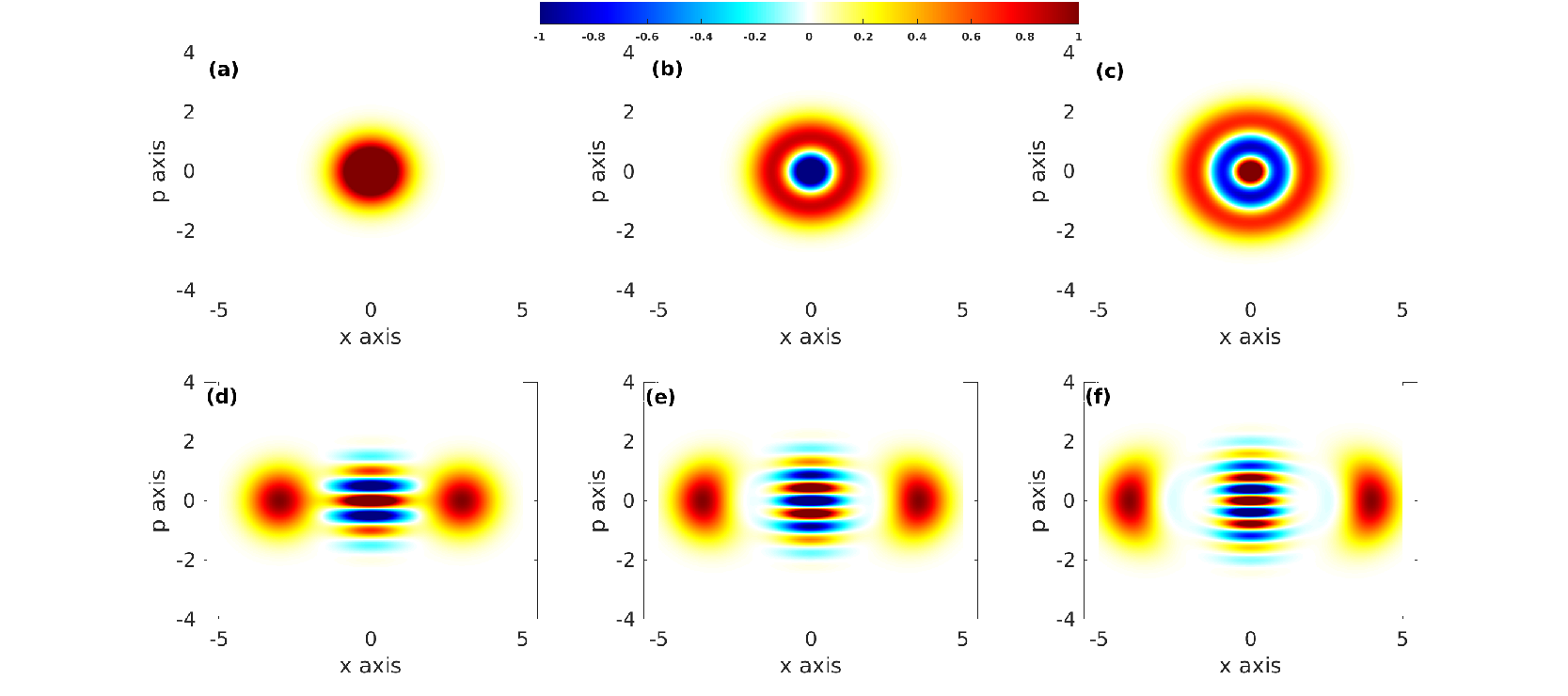}
\caption{(Color online) 
 Effect of photon addition on even cat  state (a) results in change in parity of the Wigner function at the origin (b), which is restored after addition of the second photon (c) for $\alpha = 0.3 $. For larger value of $\alpha$ ($\alpha =3$), the Gaussians are well separated (d), with a $\pi$ phase shift at origin after addition of a single photon (e). The addition of another photon leads to  fringe shift of two units, with a more pronounced squeezing (f) as compared to (e).}
\label{ssphase0}
\end{figure}
\end{center}

\twocolumngrid

Two photon added cat state takes the form:  
$$\psi_{(cat,2)}(x)  =  A \{ \psi_{(\alpha,2)}(x)+ \psi_{(-\alpha,2)}(x) \} $$

where,
$$\psi_{(\alpha,2)}(x)  =\left[\sqrt{2}\psi_{2}(x-\alpha_{1})+\sqrt{2}\alpha^{*}\psi_{1}(x-\alpha_{1})\right.$$
\begin{equation}
~~~~~~~\left.+({\alpha^{*}}^{2}/2)\psi_{0}(x-\alpha_{1})\right]e^{i\alpha_{2}(x-\alpha_{1}/2)}
\end{equation}

with

$$A=\sqrt{\frac{2}{\sqrt{\pi} (|\alpha|^4+8+8|\alpha|^2)+ (|\alpha|^4+8-8|\alpha|^2)e^{-|\alpha|^2}}}.$$
Akin to the SPACS, Wigner function for the present case can also be represented as a combination of Guassians and interference terms. Fig.\ref{ssphase0}(a) depicts the Wigner function for $\alpha =0.3$, where the two constituent coherent states have strong overlap. Evidently, for this \textit{faint} \cite{faintlaser} wave packet, although parameter $\alpha$ is small, the change in parity of the Wigner function makes the photon addition measurable \cite{tomo-parity}, as addition of a single photon  to the even cat state leads to negativity of the Wigner function. Addition of one more photon makes  the Wigner function positive at the origin, with zeros (white circle) of Wigner function separating the visible positive and negative domains in phase space as seen in Fig.\ref{ssphase0}(c).

 The larger values of $\alpha$ lead to the manifestation of the phase space interference as  clearly seen in Fig.\ref{ssphase0}(d), for which $\alpha =3$. In the process, the constituent Gaussians get well separated. Addition of a single photon to this even cat state leads to a $\pi$ phase shift at the origin, where the Wigner function displays negativity as depicted in Fig.\ref{ssphase0}(e). Squeezing of the Gaussians along X quadrature is also evident \cite{agarwal}.

Addition of another photon shifts the fringes by one more unit and restores positivity at the origin, as depicted in Fig.\ref{ssphase0}(f). Squeezing along X quadrature is more prominent, making it closer to number state \cite{agarwal}. 
After addition of two photons, Wigner function acquires two additional rings of zeros in comparison to original cat state.

It is worth emphasizing that photon added cat state is orthogonal to cat state itself, making it distinguishable.
Photon addition leads to sub-Fourier shift in the cat state as is evident from interference. More specifically, the shift at origin is given by $\pi/(2\alpha_{1})$ as has been  experimentally demonstrated by frequency-resolved optical grating (FROG) \cite{praxemayer}.

\section{PHOTON ADDED KITTEN STATE}
We now investigate the changes in interference pattern of kitten state by addition of a single photon. 

 \begin{figure}[H]
\includegraphics[width=8.50 cm,height=6.0 cm]{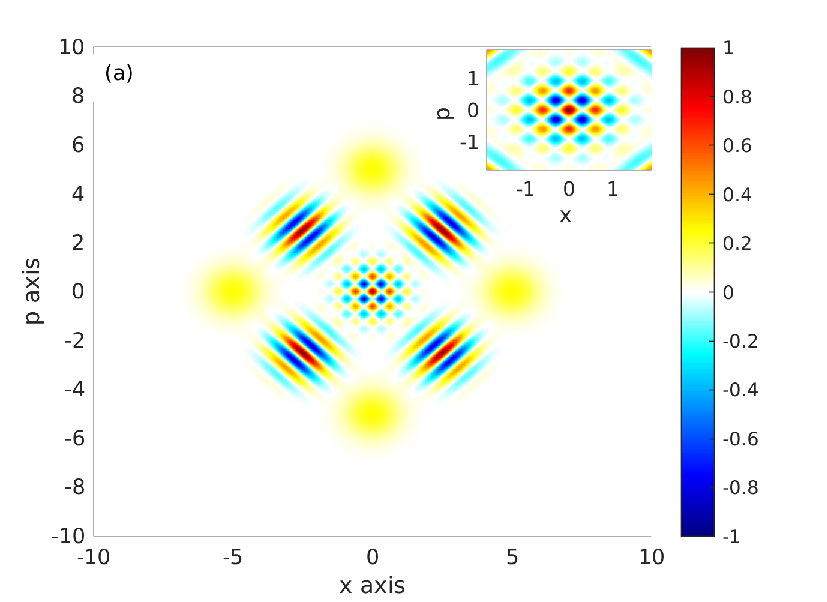}

\includegraphics[width=8.50 cm,height=6.0 cm]{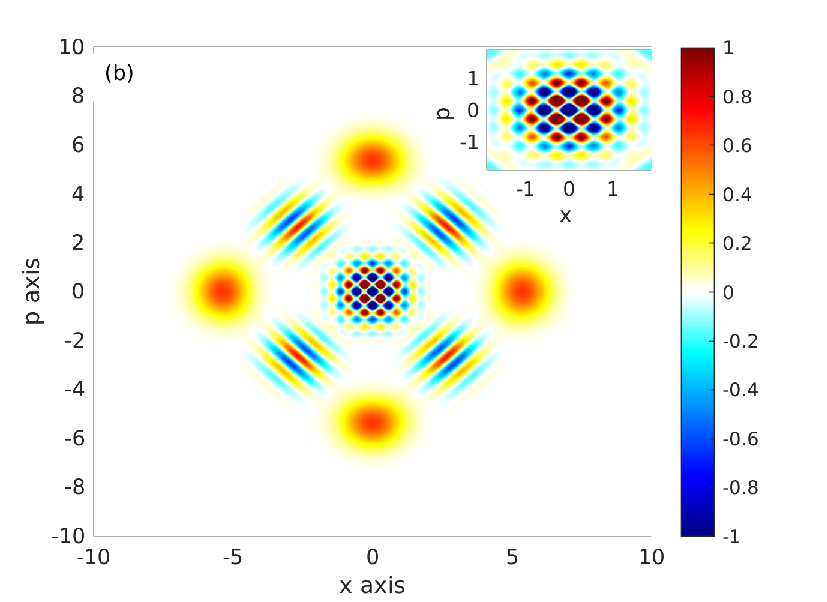}
\centering
\caption{(Color online) Addition of a photon to the kitten state (a) shows  a diagonal shift of the sub-Planck tiles, with the positions of maxima and minima getting interchanged (b). The corresponding insets clearly depict the shift of the tiles and the stronger visibility of interference after photon addition.}

\label{compass1}
\end{figure}

As pointed out earlier, the kitten state shows sub-Planck structures in phase space. Fig.\ref{ssphase0}(a) depicts these sub-Planck scale patches of interoferometric origin, with well defined tiles displayed in the inset.

 These tiles get diagonally shifted after photon addition with their  maxima and minima positions getting interchanged. This is clearly seen in Fig.\ref{compass1}(b), with the inset highlighting greater visibility after a photon addition.

\section{Photon statistics of general cat state under photon addition}

The general cat state (Eq.\ref{cat0}) is characterized by super-Poissonian and sub-Poissonian statistics for $\theta<\pi/2$ and $\theta>\pi/2$, respectively. Interestingly, the Yurke-Stoler state, with $\theta =\pi/2$, displays Poissonian behavior \cite{gerry}.

\begin{figure}[H]
    \centering
    \includegraphics[width=8.6 cm,height=6.0 cm]{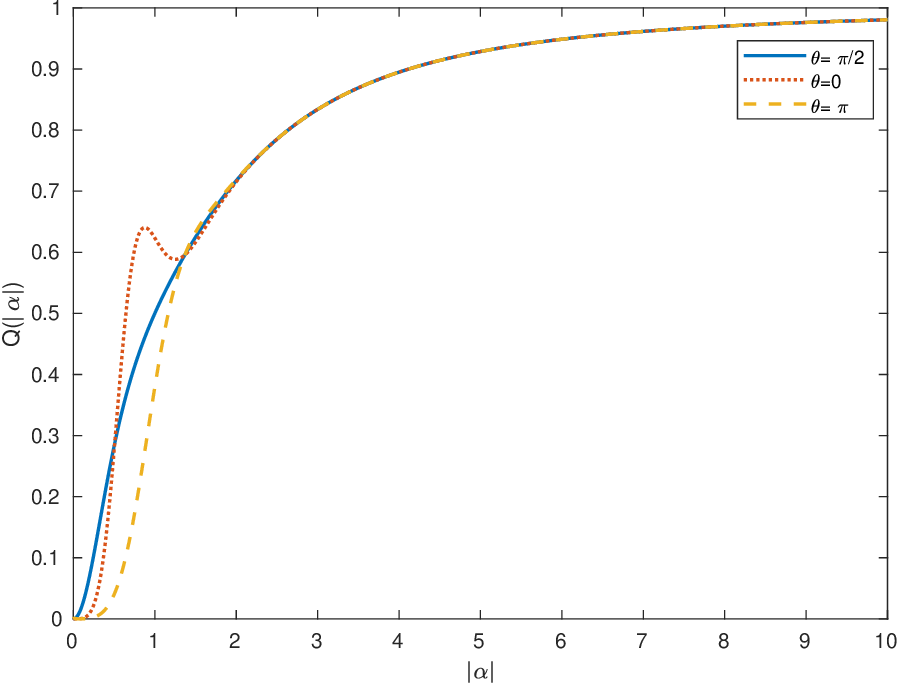}
   
    \caption{ Q parameter plotted as a function of absolute of displacement ($\alpha$) for different photon added cat states: Yurke-Stoler state ($\theta=\pi/2$), even cat state ($\theta=0$) and odd cat state ($\theta=\pi$), showing non-classicality ($Q <1$) for all values of $\theta$.}
    \label{fig:ys}
\end{figure}

The Poissonian statistics of one photon added cat state is studied using Q parameter \cite{agarwal}:
 $$ Q=\frac{\langle n^2 \rangle-\langle n \rangle^2}{ \langle n \rangle}.$$
 For photon added general cat state ($0\leq\theta\leq\pi$), one finds  
    

    
$$\,\langle n^2 \rangle = \frac{B_{-}(|\alpha|^6 + 7|\alpha|^2) + B_{+}(6|\alpha|^{4} + 1)}{|\alpha|^2 B_{-} + B_{+}}$$ 

and
$$\langle n \rangle =  \frac{B_{+}(|\alpha|^4  +1) + 3|\alpha|^{2}B_{-}}{|\alpha|^2 B_{-}  + B_{+}},$$

with $$B_{\mathbf{\pm}} = 1\pm e^{-2|\alpha|^{2}}cos\theta.$$
Remarkably, addition of a single photon to  the general cat state changes its statistical behavior to sub-Poissonian for  $\theta\leq\pi/2$. However, after photon addition, statistics of this state remains sub-Poissonian for $\theta$  ranging from $\pi/2$ to $\pi$ .  Fig.\ref{fig:ys} displays Q parameter as a function of  $|\alpha|$. It is evident that after a photon addition, Q parameter is less than one, for $\theta=\pi/2$ (YS state) and $\theta < \pi/2$ .

\section{Conclusion}

In conclusion, addition of single photon to the cat state led to observable parity change of the Wigner function, in the process causing a negative phase space domain controlled by average photon number $|\alpha|^2$ of constituent coherent states. Interference in phase space reveals fringe shifts by one and two units, after addition of one and two photons, respectively. After photon addition, Wigner function also displays squeezing  of the individual Gaussians, comprising the cat state. The fundamental tiles get shifted diagonally with interchanged positions of maxima and minima in the central interference pattern of the kitten state after a photon addition.
Statistics of Yurke-Stoler state changes from Poissionian to sub-Poissionian after one photon addition.

Changes in cat and kitten states after photon addition can find potential use in photon detection and precision metrology \cite{tetra,agarwalpathak}.
The cat and kitten states are easily destabilized by environmental induced decoherence effects as small changes in quadrature variables make them orthogonal \cite{agarwalpathak, asmitapan}. Recently, sub-Planck structures have been demonstrated for quantum hypercube states which are robust against noise \cite{hypercube}. This state may be potentially useful for single photon sensing. It is worth noting that sub-Planck structure and weak value measurement are related, since both of them have interferometric origin \cite{weakpan}. The efficacy of weak measurement for sensing  photon induced effect is currently under progress and will be reported elsewhere.

\section{Acknowledgement}

Arman is thankful to the University Grants Commission and Council of Scientific and Industrial Research, New Delhi, Government of India for Junior Research Fellowship at IISER Kolkata. GT is thankful to IISER Kolkata for hospitality and financial support from the Interdisciplinary Cyber Physical Systems (ICPS) program of the Department of Science and Technology (DST), India through Grant No.: DST/ICPS/QuEST/Theme-1/2019/6.

\bibliography{prxformat.bib}

\end{document}